\documentclass{iau}
\usepackage{graphicx}

\title[IAUS291.~~Central compact objects]{Central compact objects and \\ their
magnetic fields}

\author[W.~C.~G.~Ho]
{Wynn C.~G.~Ho}

\affiliation{School of Mathematics, University of Southampton, Southampton, SO17 1BJ, UK \\ email: {\tt wynnho@slac.stanford.edu}}

\pubyear{2012}
\volume{291}  
\jname{\mbox{Neutron Stars and Pulsars: Challenges and Opportunities after 80 years}}
\editors{J.~van Leeuwen, ed.}
\begin{document}

\maketitle

\begin{abstract}
Central compact objects (CCOs) are neutron stars that are found near the
center of supernova remnants, and their association with supernova remnants
indicates these neutron stars are young ($\lesssim 10^4\mbox{ yr}$).
Here we review the observational properties of CCOs and discuss implications,
especially their inferred magnetic fields.
X-ray timing and spectral measurements suggest CCOs have relatively
weak surface magnetic fields ($\sim 10^{10}-10^{11}\mbox{ G}$).
We argue that, rather than being created with intrinsically weak fields,
CCOs are born with strong fields and we are only seeing a weak surface field
that is transitory and evolving.
This could imply that CCOs are one manifestation in a unified
picture of neutron stars.
\keywords{pulsars: general, stars: magnetic field, stars: neutron,
supernova remnants}
\end{abstract}

\section{Introduction} \label{sec:intro}
There are a variety of manifestations/classifications of neutron stars.
Here we discuss the class dubbed central compact objects (CCOs).
CCOs are very loosely defined but are generally characterized by
the following observed properties:
(1) CCOs are associated with supernova remnants (SNRs) and are therefore
young (with ages $<\mbox{a few}\times 10^4\mbox{ yr}$),
(2) CCOs possess thermal X-ray flux that is relatively constant
(with X-ray luminosity $L_{\mathrm X}\sim 10^{33}\mbox{ ergs s$^{-1}$}$
and a spectrum that can be fit by blackbodies from small hot emitting areas),
and (3) CCOs have no optical or radio counterpart or pulsar wind nebula
(see \cite[De Luca 2008]{deluca08};
\cite[Gotthelf \& Halpern 2008]{gotthelfhalpern08}, for observational review,
including other CCOs not discussed here;
see also \cite[Halpern \& Gotthelf 2010]{halperngotthelf10}).

Only three CCOs currently have a spin period $P$ measured (as well as a
measurement or upper limit on the time derivative of spin period $\dot{P}$):
PSR~J0821$-$4300 in SNR Puppis~A
has $P=0.112\mbox{ s}$ and $\dot{P}<3.5\times 10^{-16}\mbox{ s s$^{-1}$}$
(\cite[Gotthelf et al. 2010]{gotthelfetal10}),
1E~1207.4$-$5209 in SNR PKS~1209$-$51/52 (also known as G296.5+10.0)
has two comparable timing solutions with $P=0.424\mbox{ s}$ and
$\dot{P}=2.13\times 10^{-17}\mbox{ s s$^{-1}$}$ or
$1.26\times 10^{-16}\mbox{ s s$^{-1}$}$
(\cite[Halpern \& Gotthelf 2011]{halperngotthelf11}), and
PSR~J1852+0040 in SNR Kesteven~79 has
$P=0.105\mbox{ s}$ and $\dot{P}=8.68\times 10^{-18}\mbox{ s s$^{-1}$}$
(\cite[Halpern \& Gotthelf 2010]{halperngotthelf10}).
We hereafter refer to these three CCOs as Puppis~A, 1E~1207, and Kes~79,
respectively, and only discuss them since we are primarily
interested in their magnetic fields.

\begin{figure}[t]
\begin{center}
\includegraphics[width=3.2in]{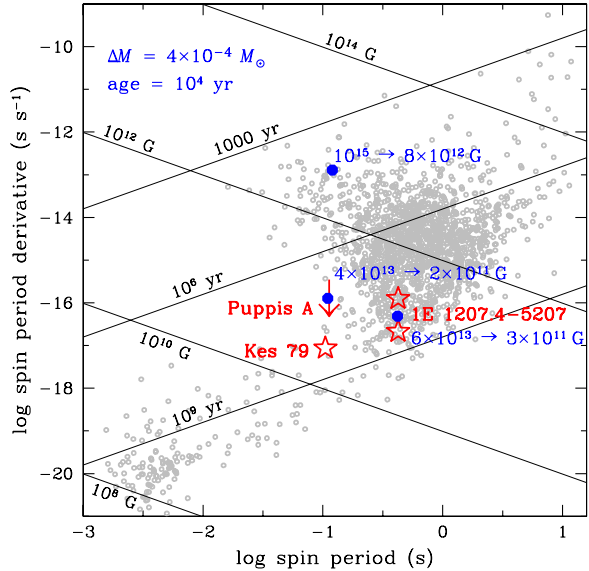}
 \caption{
Pulsar spin period $P$ versus spin period time derivative $\dot{P}$.
Small open circles are observed values taken from the ATNF Pulsar Catalogue
(\cite[Manchester et al. 2005]{manchesteretal05}).
Diagonal lines indicate constant characteristic age ($=P/2\dot{P}$) and
inferred magnetic field [$=3.2\times10^{19}\mbox{ G }(P\dot{P})^{1/2}$].
Stars (or arrow for upper limit on $\dot{P}$) denote CCOs at their
observed values (1E~1207 has two comparable solutions for $\dot{P}$).
Large closed circles indicate where these CCOs would be at age
$=10^4\mbox{ yr}$ if their initial, crust-confined magnetic field
(given by the left number) is submerged by an accreted mass
$\Delta M/M_\odot=4\times 10^{-4}$;
right number is the surface magnetic field at $10^4\mbox{ yr}$.
}
   \label{fig:ppdot}
\end{center}
\end{figure}

The spin period derivative values for CCOs are low compared to most radio
pulsars (see Fig.~\ref{fig:ppdot}).
Assuming their current $\dot{P}$ is a historical maximum or constant
(which may not necessarily be true; see, e.g.,
\cite[Muslimov \& Page 1996]{muslimovpage96};
\cite[Geppert et al. 1999]{geppertetal99};
\cite[Ho 2011]{ho11};
\cite[Ho \& Andersson 2012]{hoandersson12};
\cite[Pons et al. 2012]{ponsetal12}), then
(1) their current spin period is approximately their spin period at birth,
(2) their characteristic age $\tau_{\mathrm{c}}$ $(=P/2\dot{P})\gg$~true~age,
where the true age of
Puppis~A is $4450 \pm 750\mbox{ yr}$
(\cite[Becker et al. 2012]{beckeretal12}),
1E~1207 is $7\mbox{ kyr}$ with a factor of 3 uncertainty
(\cite[Roger et al. 1988]{rogeretal88}), and
Kes~79 is $5.4-7.5\mbox{ kyr}$ (\cite[Sun et al. 2004]{sunetal04}),
(3) their X-ray luminosity cannot be powered by rotational energy loss
since $L_{\mathrm X}>\dot{E}=4\pi^2I\dot{P}/P^3$
(which could explain their non-detection in radio;
see, e.g., \cite[Ho 2012]{ho12}), and
(4) they possess weak magnetic fields $B\sim 10^{10}-10^{11}\mbox{ G}$
(see next).

\section{Magnetic field of CCOs} \label{sec:magb}
There are two primary methods for determining neutron star magnetic fields.
The first involves timing observations, i.e., measuring neutron
star spin period $P$ and period derivative $\dot{P}$.
Assuming that the rotational energy of the pulsar decreases as a result
of the emission of magnetic dipole radiation, the field at the
magnetic equator $B_{\mathrm{e}}$ can be inferred from $P$ and $\dot{P}$,
i.e., $B_{\mathrm{e}}=3.2\times 10^{19}\mbox{ G }(P\dot{P})^{1/2}$;
note that numerical calculations of pulsar magnetospheres yields
$B_{\mathrm{e}}\approx 2.6\times 10^{19}\mbox{ G }
[P\dot{P}/(1+\sin^2\alpha)]^{1/2}$,
where $\alpha$ is the angle between the rotation and magnetic axes
(\cite[Spitkovsky 2006]{spitkovsky06}).
Using their measured values of $P$ and $\dot{P}$, CCOs have an inferred
magnetic field $B \sim 10^{10}-10^{11}\mbox{ G}$ (see Fig.~\ref{fig:ppdot}).

The second method involves spectral measurements, i.e., identifying features
in the neutron star spectrum with particular magnetic processes.
Puppis~A has a possible emission line at 0.7$-$0.8~keV
(\cite[Gotthelf \& Halpern 2009]{gotthelfhalpern09};
\cite[De Luca et al. 2012]{delucaetal12}),
and 1E~1207 has broad absorption lines at 0.7 and 1.4~keV
(\cite[Sanwal et al. 2002]{sanwaletal02};
\cite[Bignami et al. 2003]{bignamietal03}).
If we assume that a spectral line at energy $E$ is due to electron cyclotron
resonance, then the magnetic field is
$B=10^{11}\mbox{ G }(E/\mbox{1.16 keV})(1+z_{\mathrm{g}})$,
where $1+z_{\mathrm{g}}=(1-2GM/c^2R)^{-1/2}$ is the gravitational redshift
for a neutron star of mass $M$ and radius $R$.
The observed lines suggest that CCOs have $B\sim (7-9)\times 10^{10}\mbox{ G}$,
in agreement with the fields inferred from timing measurements
(spectral fits of another CCO, the 330-yr-old neutron star in SNR
Cassiopeia~A, suggest it has $B<10^{11}$~G;
\cite[Ho \& Heinke 2009]{hoheinke09}).

The magnetic field of CCOs are in contrast to the majority of neutron
stars, which possess $B\approx 10^{12}-10^{13}\mbox{ G}$, as can been seen
from Fig.~\ref{fig:ppdot}.
Furthermore, from population synthesis studies, neutron star magnetic fields
at birth follow a lognormal distribution with an average and width $\sigma$ of
$\log B=12.95\pm0.55$ in the case of no field decay
(\cite[Faucher-Gigu\`{e}re \& Kaspi 2006]{fauchergiguerekaspi06})
and $\log B=13.25\pm0.6$ when accounting for (model-dependent) field decay
(\cite[Popov et al. 2010]{popovetal10}).
The natural question is then one of creation versus evolution: Are CCOs born
with weak fields or are CCOs born with strong fields but evolve in such a way
that they appear to have weak fields at an age of $\lesssim 10^4\mbox{ yr}$?

\cite[Halpern et al. (2007)]{halpernetal07}
argue for the former and propose
that CCOs are neutron stars that are born spinning slowly.
Because of their slow rotation, the dynamo mechanism for magnetic field
generation is ineffective, and as a result, CCOs possess weak fields
($< 10^{11}\mbox{ G}$).
However, there appears to be several problems with this creation scenario.
First, CCOs do not spin particularly slowly,
as illustrated in Fig.~\ref{fig:ppdot};
this is supported by population synthesis work, which yields a normal
distribution for neutron star spin periods at birth with an average and
width of $P=0.30\pm 0.15\mbox{ s}$ in the case of no field decay
(\cite[Faucher-Gigu\`{e}re \& Kaspi 2006]{fauchergiguerekaspi06})
and $P=0.25\pm 0.1\mbox{ s}$ when accounting for field decay
(\cite[Popov et al. 2010]{popovetal10}).
Second, a birth field $<10^{11}\mbox{ G}$ would require CCOs to be
$\gtrsim 4\sigma$
from the peak of the neutron star distribution, and therefore there should be
very few of them relative to the normal pulsar population.
But this is counter to their observed numbers.
For example, \cite[De Luca (2008)]{deluca08} finds six CCOs, compared to
fourteen radio pulsars, in all known SNRs within 5~kpc
(see also \cite[Halpern \& Gotthelf 2010]{halperngotthelf10}),
and \cite[Kaspi (2010)]{kaspi10} estimates a CCO birthrate of
$\sim 0.0004\mbox{ yr$^{-1}$}$ (since all known CCOs are
$\lesssim 7\mbox{ kyr}$ old) and $\gtrsim 10^6$ CCOs in the Galaxy
(comparable to the number of strong magnetic field neutron stars).
We consider now an alternative to the creation scenario, namely evolution.

\section{Modeling magnetic field evolution} \label{sec:evolb}
In the evolution scenario, CCOs are born with strong fields
($B>10^{12}\mbox{ G}$), and these fields either decayed rapidly to their
current strengths or were buried by an early episode of accretion and
are emerging or emerged recently.
Magnetic field diffusion and decay conventionally occurs on the Ohmic timescale
$\tau_{\mathrm{Ohm}}=4\pi\sigma_{\mathrm{c}}L^2/c^2 \sim 4\times 10^5\mbox{ yr }
(\sigma_{\mathrm{c}}/10^{24}\mbox{ s$^{-1}$})(L/1\mbox{ km})^2$,
where $\sigma_{\mathrm{c}}$ is the electrical conductivity, $L$ is the
lengthscale over which decay occurs, and 1~km is the approximate size of
the stellar crust
(see \cite[Goldreich \& Reisenegger 1992]{goldreichreisenegger92}).
Thus fast decay from $\sim 10^{13}\mbox{ G}$ to $\sim 10^{11}\mbox{ G}$
could only have occurred in CCOs if the field is confined to very shallow
layers in the star (\cite[Ho 2011]{ho11}).

In \cite[Ho (2011)]{ho11}, we compare the observed properties of Puppis~A,
1E~1207, and Kes~79 to our calculations of the evolution of a buried magnetic
field.
We assume the field is buried deep beneath the surface
(\cite[Romani 1990]{romani90}), perhaps by a post-supernova episode of
hypercritical accretion (\cite[Chevalier 1989]{chevalier89};
\cite[Geppert et al. 1999]{geppertetal99};
\cite[Bernal et al. 2010]{bernaletal10}).
These fields then diffuse to the surface on the timescale of
$10^2-10^4\mbox{ yr}$, so that only now do we see a surface field
$\sim 10^{10}-10^{11}\mbox{ G}$.
We solve the induction equation, $\partial B/\partial t
 = -\nabla\times[(c^2/4\pi\sigma_{\mathrm{c}})\nabla\times B]
\sim B/\tau_{\mathrm{Ohm}}$, in one spatial dimension
(see also \cite[Muslimov \& Page 1995]{muslimovpage95};
\cite[Geppert et al. 1999]{geppertetal99}, for non-CCOs),
while \cite[Vigan\`{o} \& Pons (2012)]{viganopons12} perform two-dimensional
simulations (and thus are able to account for Hall drift) of burial and
emergence of magnetic fields in CCOs.
Fig.~\ref{fig:ppdot} shows examples of how evolution of an initially
submerged magnetic field could change $P$ and $\dot{P}$ for the CCOs.
For a field that is confined to the crust, we also find a unique relationship
between accreted mass $\Delta M$ and birth magnetic field, with a minimum
$\log B\approx 11.4-11.7$.
We find that measuring $dB/dt$ or the pulsar braking index would allow a
determination of $\Delta M$, $B$, and the field configuration, e.g.,
the field is purely in the crust if $dB/dt<0$,
while $\Delta M$ is large and the field is buried deep if $dB/dt>0$ and large.
We note that the (candidate) emission line seen in Puppis~A has decreased in
energy by 10\% in 8.5~yr (\cite[De Luca et al. 2012]{delucaetal12}); if this
is associated with a decaying magnetic field, then a purely crustal field is
implied, although the decay may be too rapid.
We also note that optical/IR observations of 1E~1207 place a limit of
$\Delta M<10^{-6}M_\odot$ on the initial mass of a debris disk
(\cite[De Luca et al. 2011]{delucaetal11}).

\section{Modeling the magnetized atmosphere spectrum of CCOs}
\label{sec:spectra}
In addition to advancements in understanding CCO timing properties,
progress has been made in modeling their spectra.
The observed thermal radiation originates in a thin atmospheric layer
(with scale height $\sim 1\mbox{ cm}$) that covers the stellar surface.
The properties of the atmosphere, such as the magnetic field,
chemical composition, and radiative opacities, directly determine the
characteristics of the observed spectrum
(see, e.g., \cite[Zavlin 2009]{zavlin09}, for review).
Very importantly, magnetic fields
$B>e^3m_{\mathrm{e}}^2c/\hbar^3=2.35\times 10^{9}\mbox{ G}$ significantly
increase the binding energy of atoms, molecules, and other bound states,
and their abundances can be appreciable in the atmospheres of neutron stars
(see \cite[Lai 2001]{lai01}, for review).
Furthermore, when $B\sim 10^{11}\mbox{ G }(T/\mbox{a few}\times 10^6\mbox{ K})$,
models of atmosphere spectra must properly account for quantum and thermal
effects in the Gaunt factor or Coulomb logarithm, which give rise to strong
cyclotron harmonics in the opacity
(\cite[Pavlov \& Panov 1976]{pavlovpanov76};
\cite[Pavlov et al. 1980]{pavlovetal80}; \cite[Potekhin 2010]{potekhin10};
\cite[Suleimanov et al. 2010, 2012]{suleimanovetal10,suleimanovetal12}).
These effects are needed in order to interpret the strong absorption lines
seen in 1E~1207 as the result of electron cyclotron resonance in an
atmosphere with $B\sim 7\times 10^{10}\mbox{ G}$, where the observed
0.7 and 1.4~keV lines are the fundamental and first harmonic, respectively
(see, e.g., \cite[Ho \& Mori 2008]{homori08}, for alternative interpretations).

\begin{figure}[t]
\begin{center}
\includegraphics[width=\textwidth]{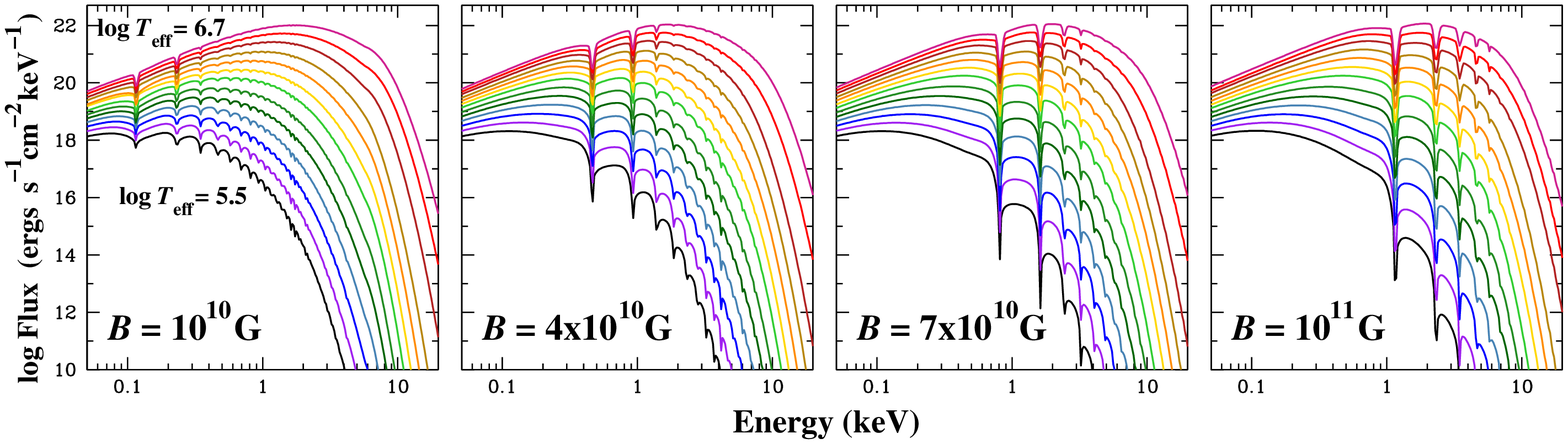}
 \caption{
Fully ionized hydrogen atmosphere model spectra for
effective temperatures $\log T_{\mathrm{eff}}=5.5-6.7$ and
magnetic fields
$B=10^{10}, 4\times 10^{10}, 7\times 10^{10}, 10^{11}\mbox{ G}$,
where the field is oriented parallel to the surface normal.
}
   \label{fig:spectra}
\end{center}
\end{figure}

\begin{figure}[t]
 \begin{minipage}[b]{0.44\textwidth}%
 \centering
 \includegraphics[angle=0,width=1\textwidth]{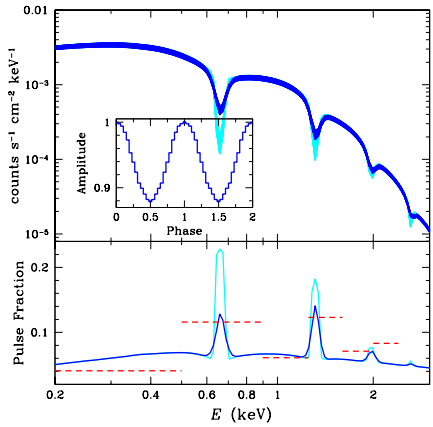}%
\end{minipage}%
\begin{minipage}[b]{0.56\textwidth}%
 \caption{
Top: Atmosphere model spectra (raw and convolved with \textit{XMM-Newton}
EPIC-pn energy resolution) at different rotation phases.
Inset: Energy-integrated (0.01-10 keV) light curve as a function of
rotation phase.  Bottom: Pulse fraction as a function of energy, where
pulse fraction
$=(F_{\mathrm{max}}-F_{\mathrm{min}})/(F_{\mathrm{max}}+F_{\mathrm{min}})$.
The dashed horizontal lines indicate the observed pulse fractions
over the given energy range for 1E~1207
(see \cite[De Luca et al. 2004]{delucaetal04}).
   \label{fig:model1207}
}
\vspace{5mm}%
\end{minipage}%
\end{figure}

We construct fully ionized hydrogen atmosphere models using the method
described in \cite[Ho \& Lai (2001)]{holai01} and using
\cite[Potekhin \& Chabrier (2003)]{potekhinchabrier03}
to calculate Gaunt factors and
\cite[Suleimanov et al. (2012)]{suleimanovetal12}
to account for thermal effects.
Examples of the resulting spectra are shown in Fig.~\ref{fig:spectra}.
We note that these weak field ($B=10^{10}-10^{11}\mbox{ G}$) neutron star
atmosphere spectra will be implemented in XSPEC under NSMAX
(\cite[Ho et al. 2008]{hoetal08}),
while partially ionized hydrogen models will be the subject of future work.
The spectra shown in Fig.~\ref{fig:spectra} only describe emission from either
a local patch of the stellar surface with a particular effective temperature
and magnetic field or a star with a uniform temperature and radial magnetic
field of uniform strength.  By taking into account surface magnetic field
and temperature distributions, we can construct more physical models of
emission from neutron stars (see \cite[Ho 2007]{ho07}, for details).
As an illustration, Fig.~\ref{fig:model1207} shows the phase-resolved
model spectra, pulse profile, and pulse fraction.
We assume here that $1+z_{\mathrm{g}}=1.235$ and
angles between rotation and magnetic axes and between rotation axis and
observer are $5^\circ$ and $25^\circ$.
The hot spot covers magnetic colatitudes $0-30^\circ$ and has
effective temperature $T_{\mathrm{eff}}=2\times 10^6\mbox{ G}$ and
magnetic field $B=7\times 10^{10}\mbox{ G}$ that is oriented parallel to
the surface normal.
The bottom panel shows the pulse fraction in different energy bands that
is measured for 1E~1207 (\cite[De Luca et al. 2004]{delucaetal04}); what
is particularly noteworthy is that the pulse fraction is larger at the
spectral lines, and accounting for thermal effects in the model appears
to be necessary to achieve these higher pulse fractions
(see also \cite[Suleimanov et al. 2012]{suleimanovetal12}).

\section{Discussion} \label{sec:discuss}
>From timing and spectral studies, CCOs appear to have relatively weak surface
magnetic fields ($B\approx 10^{10}-10^{11}\mbox{ G}$).
The question arises as to whether CCOs are born with inherently weak
magnetic fields (creation scenario)
or they are born with much stronger fields but these fields were buried
and are evolving (evolution scenario).
The creation explanation is simple, but as discussed in Sec.~\ref{sec:magb},
there are problems.
There are also problems with the evolution scenario
(see, e.g., \cite[Halpern \& Gotthelf 2010]{halperngotthelf10}).
Nevertheless, evolution of magnetic fields seems natural, and there is
evidence in favor of buried magnetic fields in CCOs.
For example, \cite[Shabaltas \& Lai (2012)]{shabaltaslai12} construct models
with strong toroidal fields ($B>10^{14}\mbox{ G}$) in the crust to explain
the high pulse fraction of Kes~79 ($64\pm2\%$;
\cite[Halpern \& Gotthelf 2010]{halperngotthelf10}).
Also \cite[Gotthelf et al. (2010)]{gotthelfetal10} argue that a
strong tangential field in the crust can explain the small hot spots seen
on Puppis~A, and this is confirmed qualitatively with magneto-thermal
simulations by \cite[Vigan\`{o} \& Pons (2012)]{viganopons12}.

If CCOs have buried magnetic fields, then this sub-surface field is likely
to be $\gtrsim 10^{12}\mbox{ G}$.  If burial is shallow, then the surface
field is currently decaying.
If burial is deep, then the surface field is growing rapidly and could lead
to a rapid change in spin parameters.
We see from Fig.~\ref{fig:ppdot} that CCOs reside in a relatively
underpopulated region of $P-\dot{P}$ parameter space
(see also \cite[Halpern \& Gotthelf 2010]{halperngotthelf10};
\cite[Kaspi 2010]{kaspi10}).
Rapid spin evolution could mean that CCOs are moving out of this region
quickly and joining the majority of the pulsar population at longer spin
periods, higher $\dot{P}$, and stronger observed magnetic fields.
Thus magnetic field evolution could facilitate the unification of
CCOs with other classes of neutron stars
(see \cite[Kaspi 2010]{kaspi10}; \cite[Popov et al. 2010]{popovetal10};
\cite[Ho 2012]{ho12}).

~\\
WCGH thanks Alexander Potekhin for discussions,
appreciates the use of the computer facilities at KIPAC,
and acknowledges support from the RAS and STFC in the UK.

\end{document}